\documentclass[12pt]{article}
\usepackage{epsf}
\usepackage{amsmath}
\usepackage{graphics}
\usepackage{cite}

\renewcommand{\thefootnote}{\#\arabic{footnote}}
\begin{document}

\newcommand{\gtrsim}{ \mathop{}_{\textstyle \sim}^{\textstyle >} }
\newcommand{\lesssim}{ \mathop{}_{\textstyle \sim}^{\textstyle <} }

\newcommand{\rem}[1]{{\bf #1}}

\renewcommand{\thefootnote}{\fnsymbol{footnote}}
\setcounter{footnote}{0}
\begin{titlepage}

\def\thefootnote{\fnsymbol{footnote}}

\begin{center}
\hfill September 2013\\
\vskip .5in
\bigskip
\bigskip
{\Large \bf Predictions of Entropic Gravitation}

\vskip .45in

{\bf Paul H. Frampton\footnote{frampton@physics.unc.edu}}

{\em Department of Physics and Astronomy,}

{\em University of North Carolina at Chapel Hill, NC 27599, USA}

\begin{center}
and
\end{center}

{\bf Gabriel Karl\footnote{gk@physics.uoguelph.ca}}

{\em Department of Physics,}

{\em University of Guelph, Guelph, Ontario N1G 2W1, Canada}

\end{center}

\vskip .4in
\begin{abstract}
We discuss consequences of an entropic view of gravity,
which differs in details from the original proposals
of Jacobson and Verlinde. We assume the entropy is localized
in the degrees of freedom of the systems interacting
gravitationally. We then find there is a density-dependent
minimum length for Newtonian gravity. This makes 
renormalizability issues irrelevant. It also follows
that at nanometer scales Newton's universal
law fails, black holes with mass 
$M_{BH} \lesssim 10^{-13} M_{\odot}$
do not exist and all Hawking radiation 
is at unobservably low temperature.
\end{abstract}
\end{titlepage}

\renewcommand{\thepage}{\arabic{page}}
\setcounter{page}{1}
\renewcommand{\thefootnote}{\#\arabic{footnote}}

\newpage

Gauge theories such as quantum electrodynamics \cite{QED}, 
constructed by Dyson,
Feynman, Schwinger and Tomonaga, quantum chromodynamics \cite{QCD} , due to
Fritzsch, Gell-Mann, Gross, Nambu and Wilczek, and
the standard model \cite{SM}, originated by Glashow, 't Hooft, Salam, Veltman
and Weinberg, are renormalizable. This despite the divergences which
occur because of employing singular products
of fields at the same spacetime point.

When the same approach is applied to gravity, one finds
non-renormalizability, because the gravitational interaction
is too large at very short distances. This is the principal reason
why string theory is favored in which the uncontrollable
divergences are ameliorated by replacing local fields by extended
objects whose interactions are suitably spread out in spacetime
and can therby avoid the previous divergences and lead to
a finite theory.

In this note we suggest that the non-renormalizability of the
non-string approach may be based on a false presumption.

In the quantum electrodynamics of photons and electrons,
there exists a primordial vertex of the photon
to the electron with coupling $e$, the electronic charge, and
the whole renormalizable theory is built up upon this.
Similarly for quantum chromodynamics there is a primordial
gluon-quark vertex.

The presumption in quantum gravity is generically
that there exists a corresponding
primordial graviton-electron vertex (for example)
with a non-zero coupling related to Newton's constant.
The non-renormalizability can then be ascribed
to the fact that this coupling
becomes too large at very  short distances, when compared to the corresponding
coupling in a renormalizable theory.

\bigskip

If we consider two electrons, at a very short distance apart,
 conventional wisdom
is that the gravitational force is more than
forty orders of magnitude smaller
than the electromagnetic force.

\bigskip

If we adopt an entropic view \cite{Jacobson,Verlinde} 
(for one application, see \cite{EFS}),
the situation is changed.

\bigskip

According to the entropic view, the force of gravity, 
Newton's law, is explicable always as an entropy increase. 
In this case, the gravitational force between two isolated
electrons whose entropy cannot increase, 
is not merely extremely small compared to the electomagnetic
force, it is zero. 

\bigskip

At least one of the two objects must have sufficiently
many subcomponents that its entropy is adjustable.
It is known that neutrons are attracted to the Earth
which contains $\sim 10^{51}$ nucleons. The question
is then how many constituents, $N$,  are necessary. 
Probably it must satisfy $N \gg \sqrt{N}$ in order that
fluctuations are sufficiently small, so that {\it e.g.}
$N \geq 100$.

\bigskip

While it is known that neutrons are attracted to
the Earth, from experiment, when both bodies are elementary, like
electrons, we expect zero gravitational force.

\bigskip

This then implies that the primordial electron-graviton vertex is
not exceptionally large at very short distances, but actually vanishes. 

Therefore there is no reason to expect gravity to be renormalizable,
or to employ string theory. In the entropic approcach, if there
were a graviton it would be a collective excitation more like
a phonon than a photon, and not an elementary particle.
Because the gravitational force between two elementary particles
is zero, there is no reason for
conventional quantum gravity to be renormalizable.

It is known from the beautiful experiment of Colella, Overhauser
and Werner\cite{Overhauser} that a single neutron is attracted to the earth.
On the other hand,  with only two elementary particles, like two electrons
(or two neutrons) with fixed positions (and spins) we expect a zero force 
in the entropic scenario. In principle, the experiment of Colella et al
\cite{Overhauser}
could be replicated with a small sample of matter (say 1000 particles) 
as the source of the gravitational field for a split neutron beam. 
One could then observe that as the number of particles diminishes in 
the source the fluctuations would destroy the gravitational potential. 
Of course in practice such an experiment would be impossibly hard 
to achieve.
 
\bigskip

\noindent
It is known that Newton's law is valid\cite{Adelberger} down to a 
distance of a few tens of microns, or $\sim 10^{-5}$ m, while we would
expect it to break down in matter of normal density before reaching
an atomic scale, $\sim 10^{-10}$ m. 
If we require one thousand (one million) 
surrounding atoms it would suggest $\sim 10^{-9}$ m ($\sim 10^{-8}$ m)
as the shortest scale at which Newton's law could become approximately
valid. We therefore can reasonably expect a deviation from the inverse-square law
of Newton to be experimentally detectable in principle, although
almost impracticably challenging\cite{Adelberger} 
in practice, at such length scales. The assumption is being made
of normals densities as is applicable to the material
of the massive objects used in the Cavendish- and Adelberger- type experiments
\footnote{In a system with nuclear density such as a neutron star
the corresponding lower limit to the validity of Newton's
universal law would be $\sim 10^5$ times smaller.}.

\bigskip
\bigskip

\noindent
{\it Black Holes}

\bigskip

\noindent
In this entropic approach there exists a fundamental length $L$
providing a cut-off and from our discussions the value is in
the range

\begin{equation}
1 nm \leq L \leq 50 \mu.
\label{L}
\end{equation}
where the upper limit is from experiment \cite{Adelberger} and
the lower limit derives from theory.

\bigskip

\noindent
Let us take the most conservative possibility, with the least
departure from general relativity, and set $L=1$ nm. 

\bigskip

\noindent
When we consider black hole\footnote{Since the details of incorporating
black holes into entropic gravity are, to say the least, incompletely
understood, the present discussion must be regarded only
as indicative because we are assuming {\it faute de mieux} a minimal
length $L$
corresponding to normal matter densities.}
solutions such as the Schwarzschild solution
\cite{Schwarzschild}, the cut-off $L$ means that we must
consider only those solutions which exist for larger scales which
means that the Schwarzschild radius $r_S$ should satisfy
$r_S \geq L$. With $L=1$ nm this implies a black hole mass $M_{BH}$
satisfying

\begin{equation}
M_{BH} \geq 10^{18} kg \sim 10^{-7} M_{\oplus} \sim 10^{-13} M_{\odot}
\label{minBH}
\end{equation}
where the mass of the Earth and Sun are $M_{\oplus} \sim 6 \times 10^{24}$kg
and $M_{\odot} \sim 2 \times 10^{30}$ kg, respectively.

\bigskip

\noindent
The existence of the high energy cut-off at $\sim 1000$ eV implies
that black holes smaller than the lower bound Eq.(\ref{minBH})
may not be possible and this is contrary to the standard lore \cite{Carr}
concerning Primordial Black Holes (PBHs). Does this disfavor or favor
the cut-off? Since no black holes which violate Eq.(\ref{minBH})
have shown up observationally, a neutral researcher could say that
our approach is favored.

\bigskip

\noindent
The radiated photons from a black hole must have wavelength
$\lambda > L = 1$ nm. This translates into a maximim energy
$E_{\gamma} \lesssim 1000$ eV. This is insufficient generically
to create an $e^+e^-$ pair, and therefore calls into question
the standard analysis \cite{Hawking} of Hawking radiation.

\bigskip

\noindent
Our approach differs in details from those of 
Jacobson \cite{Jacobson}
and Verlinde \cite{Verlinde}. With respect to \cite{Jacobson},
that author's minimal length is the Planck length. With respect to
\cite{Verlinde}, our treatment of entropy is more local 
so that equality of action and reaction, as well 
as conservation of momentum, are {\it prima facie} compromised.

\bigskip

\noindent
In summary, while we have not dared or been competent to present
a theory, alternative to general relativity, we have speculated
on how a putative theory which incorporates the entropic emergence
of gravity will likely differ in its predictions. For all large
scales, cosmological, galactic, and terrestrial, there will be
no measurable difference. At very short scales, however,
only a few times
the atomic size charaterized by the Bohr radius, there are
predicted to be very big differences:
Newton's universal gravity is predicted to fail, black holes
with such a small radius are predicted not to exist and Hawking
radiation will be all at unobservably low temperature.

\newpage

\begin{center}

{\bf Acknowledgements}

\end{center}

\bigskip
We thank F. Englert, S.L. Glashow, J.C. Taylor,
and M. Williams for useful discussions.
This work was supported in part by the
U.S. Department of Energy under Grant No. DE-FG02-06ER41418.

\end{document}